\documentclass[sigconf]{acmart}

\renewcommand\footnotetextcopyrightpermission[1]{}
\usepackage{booktabs}
\usepackage{subcaption}
\usepackage{listings}
\usepackage{xcolor}
\setcopyright{none}
\begin{document}

\title{DRCY: Agentic Hardware Design Reviews}

\author{Kyle Dumont \quad Nicholas Herbert \quad Hayder Tirmazi\footnotemark \quad Shrikanth Upadhayaya}
\affiliation{%
  \institution{\vspace{0.5em}AllSpice Inc.}
  \country{}
}
\authorsaddresses{}

\begin{abstract}
Hardware design errors discovered after fabrication require costly physical
respins that can delay products by months. Existing electronic design automation
(EDA) tools enforce structural connectivity rules. However, they cannot verify that
connections are \emph{semantically} correct with respect to component
datasheets. For example, that a symbol's pinout matches the manufacturer's
specification, or that a voltage regulator's feedback resistors produce
the intended output. We present DRCY, the first production-ready 
multi-agent LLM system that automates first-pass schematic connection 
review by autonomously fetching component datasheets, performing pin-by-pin 
analysis against extracted specifications, and posting findings as inline 
comments on design reviews. DRCY is deployed in production on AllSpice Hub, 
a collaborative hardware design platform, where it runs as a CI/CD action 
triggered on design review submissions. DRCY is used regularly by major hardware companies 
for use-cases ranging from multi-agent vehicle design to space exploration. We describe DRCY's five-agent pipeline 
architecture, its agentic datasheet retrieval system with self-evaluation, 
and its multi-run consensus mechanism for improving reliability on safety-critical analyses.
\end{abstract}




\maketitle
\renewcommand{\thefootnote}{*}
\footnotetext{Corresponding author is hayder@allspice.io. Authors are listed in alphabetical order by last name. }
\renewcommand{\thefootnote}{\arabic{footnote}}

%% ============================================================
\section{Introduction}
\label{sec:intro}

Printed circuit board (PCB) design review is the last line of defense before a
design is sent for fabrication. PCB designs may have connection errors such as a swapped pin pair, a
misrouted power rail, or a violated voltage specification. If these errors 
are missed in a review, it results in a board respin costing weeks to months of schedule and tens of
thousands of dollars per iteration. Currently, most reviews are done manually. Manual review is thorough but
labor-intensive. For example, a reviewer must cross-reference every connection against the
relevant component datasheet, verify pin assignments, check signal polarity
conventions, and confirm compliance with electrical specifications. A
moderately complex board may have hundreds of components across dozens of
schematic pages, each backed by a datasheet that can exceed 100 pages.

EDA tools provide two classes of automated checks. Electrical rule
checks (ERC) verify structural properties such as floating pins, shorted nets, and
unidirectional conflicts. Design rule checks (DRC) verify physical
layout constraints including trace widths, clearances, and drill sizes.
Circuit simulation offers a third avenue, but requires hardware engineers to build
detailed component models and specify all input stimuli. This is a time-consuming
process that scales poorly with design complexity. Crucially, all three
approaches share a fundamental limitation: They only catch issues that the
engineer has \emph{already anticipated}. Prior to an error being detected, 
an engineer has to write an ERC rule, specify a DRC
constraint, or configure a complicated simulation testbench. In addition, none of these methods can determine whether a
connection is \emph{semantically correct}. By semantic correctness, we mean checking
whether the schematic faithfully implements the component manufacturer's recommended connections. For
example, an ERC will confirm that a microcontroller's \texttt{nRESET} pin is
connected, but not that it is connected to the correct reset circuit as
specified in the datasheet.

In this paper we present DRCY\footnote{DRCY (pronounced ``DAR-SEE'') is a proper name, not an
acronym.}, a multi-agent LLM system that fills this gap. DRCY is the
first agent built for hardware design reviews. It integrates with
AllSpice Hub~\cite{allspice} which is a collaborative platform for hardware design
with native EDA file support. DRCY runs as a CI/CD action triggered when
a hardware engineer submits a design review. DRCY is in production use by
hardware teams at Fortune~500 enterprises and growth-stage startups,
analyzing designs spanning consumer audio, robotics, autonomous vehicles,
and space exploration.

% ============================================================
\section{Background}
\label{sec:background}

This section introduces the hardware design concepts that DRCY operates
on, aimed at readers more familiar with software engineering workflows.

\smallskip\noindent\textbf{Schematics and Netlists.} 
A PCB design begins as a schematic, i.e., a diagram that specifies
which electronic components are used and how their pins are
electrically connected~\cite{art_of_electronics}. Pins are the physical terminals of a component such as
power input, data output, and ground. A net is a named
electrical connection that links two or more pins together. The complete
set of nets forms the netlist~\cite{llm_eda_survey}, which is analogous to a call graph
in software. The netlist describes the connectivity structure of the design
without specifying physical geometry. A moderately complex board may
contain hundreds of components, thousands of pins, and dozens of
schematic pages.

\smallskip\noindent\textbf{Components and Datasheets.} 
Each component on a schematic, such as a microcontroller, voltage
regulator, sensor, or passive element, is identified by a
manufacturer part number (MPN). The component manufacturer
publishes a datasheet. The datasheet is a document, which can range from a couple of pages to hundreds of pages, that
specifies pin functions, absolute maximum ratings, recommended operating
conditions, and application circuits that show how the component should be
connected~\cite{art_of_electronics}. Datasheets are the ground truth for the correctness of a design. A
connection that violates a datasheet specification, for example, driving
a 1.8\,V input with a 3.3\,V signal, can cause component damage or
a silent malfunction.

\smallskip\noindent\textbf{Design Reviews.} 
Before a PCB is sent for fabrication, the design undergoes a
design review~\cite{ipc2221}. The design review process is analogous to code review in software.
A reviewer examines the schematic and cross-references each connection
against the relevant datasheets to verify correctness. Unlike software,
where a bug can often be patched after deployment, a hardware error
discovered after fabrication requires a physical board respin. A board respin is a
new manufacturing run that typically costs tens of thousands of dollars
and delays the project by weeks to months.

\smallskip\noindent\textbf{EDA Format.} 
Schematics are authored in electronic design automation (EDA)
tools~\cite{llm_eda_survey}, also referred to as electronic
computer-aided design (ECAD) tools. Unlike software, where source code is predominantly plain text,
EDA vendors often use a proprietary binary or structured format. For example,
Altium~\cite{altium_overview}
uses \texttt{.SchDoc}, Cadence OrCAD~\cite{cadence_overview} uses \texttt{.dsn}, KiCad~\cite{kicad_overview} uses
\texttt{.kicad\_sch}, and so on. This format fragmentation means that
any tool operating on schematics must support multiple parsers.
AllSpice Hub parses all these formats, and DRCY builds on this platform capability.

% ============================================================
\section{Design}
\label{sec:design}

This section describes the agent architecture of DRCY's end to end 
pipeline from a design schematic to a design review. When a design review is submitted on AllSpice Hub, 
the AllSpice Hub first parses the ECAD source files for a diverse array of common formats 
including Altium, OrCAD, KiCad, System Capture, Xpedition, or DE-HDL. After parsing these ECAD source files,
the AllSpice Hub outputs a structured JSON representation containing components, pins, nets,
and graphical annotations. This structured JSON representation is what DRCY takes 
as its input representation. For each schematic page in this JSON representation, DRCY then
proceeds through a pipeline of five main agents.

\begin{figure*}[t]
  \centering
  \includegraphics[width=\textwidth,trim=0cm 0cm 0cm 0cm, clip]{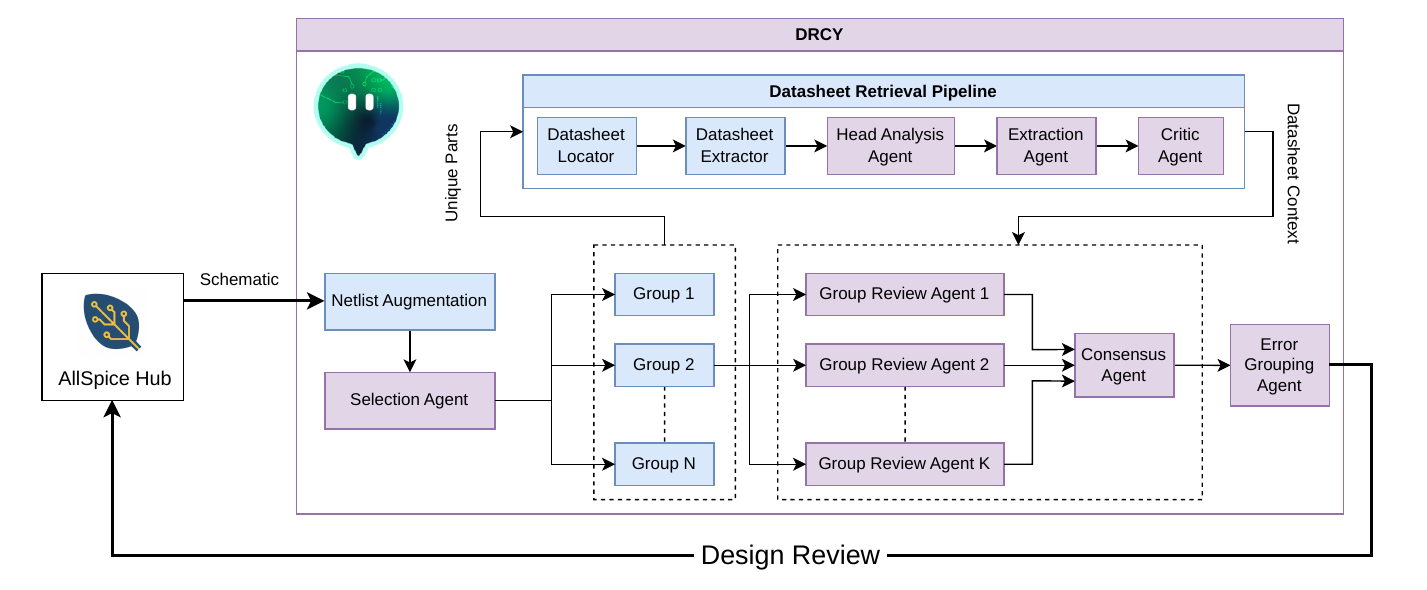}
  \caption{An overview of DRCY's architecture. A schematic received from AllSpice Hub flows through
  netlist augmentation, component selection and grouping, parallel datasheet
  retrieval, per-group review agents, consensus combination, and error
  grouping. The results from DRCY are posted back to AllSpice Hub as design review comments.}
  \label{fig:architecture}
\end{figure*}

\subsection{Netlist Augmentation}
\label{sec:netlist}

When DRCY receives structured schematic information from the AllSpice
Hub Platform, it first augments schematic data with netlist information
to provide a complete picture of electrical connectivity. The system
supports three strategies, one of which is selected automatically based
on the EDA format. For formats
with embedded net data such as KiCad, Altium, and OrCAD, nets are extracted directly
from the schematic JSON. For Cadence DE-HDL designs, a \texttt{pstxnet.dat} file is parsed.
The augmented representation is serialized
as XML for consumption by downstream agents.

\subsection{Selection}
\label{sec:pipeline}

Analyzing all components on a page in a single LLM call would exceed context
limits and dilute focus. Instead, DRCY uses a Selection Agent that takes the components in a design schematic
and partitions them into functional groups. Examples of functional groups include a voltage regulator with its feedback network and decoupling
capacitors which forms a power stage, a microcontroller with its clock source and
bypass capacitors which forms a core compute group, or an interface transceiver with its
termination resistors which forms an I/O group. For each group, the agent extracts
a manufacturer part number (MPN) or an internal part number (IPN) for every
unique component, along with any datasheet URLs present in the schematic.

\subsection{Agentic Datasheet Retrieval}
\label{sec:datasheets}

A distinguishing capability of DRCY is its fully autonomous datasheet
processing pipeline. Given a part number from the selection stage, DRCY
first uses a custom Datasheet Locator system. The Datasheet
Locator queries a configurable set of component libraries including an
API for datasheet information provided by DigiKey, custom JSON template
APIs provided by clients, custom CSV lookup tables, or files uploaded by
clients directly on AllSpice Hub. Multiple libraries can be
configured per deployment to support both standard and proprietary
components.

After a datasheet is located, the location is sent to a Datasheet
Extractor, a Playwright-based browser automation layer
that fetches distributor pages. Downloads are deduplicated so that
concurrent requests for the same part number share a single in-flight
fetch.

A Head Analysis Agent examines extracted datasheets including the PDF's
table of contents or page thumbnails to select pages containing pin
descriptions, electrical specifications, and application circuits. This
head analysis reduces the token cost of the subsequent steps in the
agentic pipeline.

After head analysis, DRCY uses an Extraction Agent to process the
selected pages and produce a compact XML representation of the
datasheet. This compact representation typically contains pin functions
with relevant metadata, absolute maximum ratings, recommended operating
conditions, internal block descriptions, and application circuit
guidance.

Finally, DRCY uses a Critic Agent to score each datasheet extraction on
a 10 point scale across four weighted dimensions: feature completeness
(25\%), pin function coverage (40\%), application information (20\%),
and typical application circuits (15\%). If the score falls below a
configurable threshold and alternative URLs remain in the queue, the
datasheet retrieval pipeline retries with a different source. The
datasheet retrieval pipeline makes up to five attempts for each
datasheet. The best-scoring extraction out of these attempts is
returned. Extracted specifications are cached with a 7-day TTL on
persistent cloud storage keyed by part number and source URL. Cache hits
bypass the entire datasheet pipeline.

\subsection{Group Review and Multi-Run Consensus}
\label{sec:review}

Group Review Agents perform the core analysis for DRCY's design review.
LLM outputs are nondeterministic, so the same prompt may yield
different findings across invocations. Rather than treating this as a liability,
DRCY exploits it as a reliability mechanism. $k$ separate instances of
a Group Review Agent are invoked for each functional group identified by
the Selection Agent (Section~\ref{sec:pipeline}). All $k$ runs execute
concurrently so latency scales sub-linearly with $k$. The Group Review
Agent receives 1) the netlist-augmented schematic, 2) the extracted
datasheet specifications for all components in the group, and 3) a
checklist for the group review. Each Group Review Agent outputs
structured JSON, mapping each component designator to a
Component Analysis containing per-pin correctness verdicts. Pins
sharing a root cause (e.g., a swapped pin pair) can be grouped under a
single analysis entry keyed by multiple pin designators
(e.g., ``1, 3'').

The $k$ Group Review Agents that ran for each Component Group are
consolidated by a separate Multi-Run Consensus Agent. A separate
Multi-Run Consensus Agent is invoked for each Component Group. The
Multi-Run Consensus Agent reconciles the results by 1) retaining
findings present in multiple runs with higher confidence, 2) critically
evaluating findings from a single run against the schematic and the
datasheets before including it, and 3) re-examining contradictory
results with full context. The Multi-Run Consensus Agent uses a
dedicated model which is configurable separately from the model used by
the Group Review agents. Note that the Multi-Run Consensus Agent has
access to the same schematic and datasheet context as the individual
Group Review Agents.

% \begin{enumerate}
%   \item Enumerate every pin on each assigned component.
%   \item Trace the net connections through the schematic, identifying what each
%         pin is connected to.
%   \item Match pin names between the schematic and datasheet, accounting for
%         naming convention differences.
%   \item Analyze each connection against the datasheet specification and produce
%         a verdict (\textsc{correct} or \textsc{incorrect}) with evidence.
%   \item For text notes on the schematic (e.g., ``DNP R3 for 1.8V output''),
%         apply a three-step classification: direct component reference, spatial
%         proximity, and semantic relevance filtering.
% \end{enumerate}
% \paragraph{Active-low signal normalization.}
% Electrical schematics encode active-low signals using diverse conventions:
% KiCad uses \texttt{\textasciitilde\{RESET\}}, Altium uses backslash-delimited
% characters, and DE-HDL uses trailing asterisks. DRCY normalizes all formats
% into Unicode combining overlines
% ($\overline{\text{R}}\overline{\text{E}}\overline{\text{S}}\overline{\text{E}}\overline{\text{T}}$)
% before analysis. The review agent's system prompt explicitly maps this
% representation to common datasheet conventions (\texttt{nRESET},
% \texttt{/RESET}, \texttt{RESET\_B}, \texttt{RESET\#}), preventing false
% positives from notation mismatches across EDA tools. 

\subsection{Error Grouping}
\label{sec:grouping}

A single root cause often manifests across multiple pins and components. For
instance, a swapped pin pair on an IC affects both the IC's pins and any
downstream components connected to the affected nets. The Error Grouping
agent identifies these relationships and assigns them deterministic group identifiers. 
These identifiers drive the comment generation for DRCY's design review. 
DRCY consolidates related errors into a single review comment with per-component detail,
instead of scattering them across multiple comments, which would imply they are independent findings.

%% ============================================================
\section{Deployment}
\label{sec:deployment}

DRCY is deployed as a Docker container that runs as a CI/CD action on AllSpice Hub. When a design review is submitted, the AllSpice Hub triggers the action,
which fetches the schematic data, runs the analysis pipeline, and posts results
back as inline review comments with Markdown-formatted tables showing per-pin
verdicts. Each comment includes a link to the datasheet consulted, and
schematic snippets are rendered as SVG overlays highlighting the relevant
portion of the page.

The system supports two operational modes. In \emph{design review mode} only modified 
schematic pages are analyzed, reducing cost and
latency for iterative reviews. In \emph{full analysis mode}, all pages are
analyzed. This is useful for initial design review or comprehensive audits. A
configurable time limit allows operators to cap execution for cost control,
with the system posting partial results for any pages completed before the
deadline.

DRCY uses a tiered model strategy to balance quality and cost. A
strong model handles the review and combination agents, where reasoning quality directly affects
accuracy. A weak model handles datasheet extraction and head analysis, where the task is more mechanical. 
By default, DRCY uses Anthropic models of different strengths as its strong and weak models.
However, custom OpenAI-compatible providers can be configured for on-premises or alternative
model deployments.

DRCY is in production use by hardware design teams at multiple Fortune~500
enterprises, a leading autonomous vehicle technology company, a commercial
space exploration company, and several growth-stage hardware startups. The system analyzes designs across Altium, OrCAD, KiCad,
System Capture, Xpedition, and DE-HDL. The system operates within a SOC~2 Type~II certified environment with
GDPR compliance. Note that customer design data is \textbf{not} used for model training. Customers requiring full data sovereignty
can deploy DRCY on-premises with their own LLM infrastructure. A typical 10-page schematic with 50 to 100 unique components completes analysis
in less than 20 minutes. OpenTelemetry tracing is also integrated throughout the pipeline,
providing per-agent latency, token consumption, and datasheet cache hit rates
for operational monitoring.

%% ============================================================
\section{Demonstration}
\label{sec:demo}

The live demonstration gives attendees a hands-on look at DRCY's
end-to-end workflow on a real hardware design containing intentionally
introduced connection errors. The presenter opens a schematic design 
on AllSpice Hub with DRCY enabled. On the schematic design, the presenter 
submits a new design review, which automatically triggers DRCY as a CI/CD action\footnote{Demo video: \url{https://www.loom.com/share/9dff06ea27b2413cbd5e50c542a08ff3}}.

\smallskip\noindent\textbf{Pipeline Walkthrough.} After the presenter submits the design review, the attendees can observe DRCY's progress
in real time through status updates on the design review. The status updates
for DRCY's progress include a progress bar and per-page status
indicators. As the pipeline for the DRCY design review executes, 
the presenter narrates each stage. The stages include, as discussed in 
Section~\ref{sec:design}, 1) the Selection Agent partitioning components
into functional groups, 2) the Datasheet Retrieval pipeline autonomously 
locating and extracting specifications from many datasheets with variable 
page counts, 3) the parallel
Group Review Agents analyzing each functional group, 4) the Multi-Run 
Consensus Agent combining the individual reviews of the Group Review Agents for each
group, and finally 5) the Error Grouping Agent deciding the granularity with which to 
report the errors detected and analyzed from DRCY's design review.

\smallskip\noindent\textbf{DRCY Review Results.} After the entire pipeline for DRCY's design review completes, 
DRCY posts inline comments directly on the design review. The presenter 
shows these inline comments which include an identification of each error that 
DRCY found along with DRCY's reasoning. More concretely, each DRCY comment includes a
table of per-pin verdicts, a detailed reasoning chain
referencing specific datasheet pins and schematic nets, a link to the
consulted datasheet, and an SVG overlay highlighting the affected region
of the schematic page.

%% ============================================================
\section{Related Work}
\label{sec:related}

Commercial EDA suites including Cadence
OrCAD~\cite{cadence_erc,cadence_drc} and
Altium~\cite{altium_erc,altium_drc} provide ERC and DRC but are limited
to structural and physical constraints. They cannot verify
datasheet-level semantic correctness, which is DRCY's focus.

Startup tools have begun to address datasheet-level verification. Cady
Solutions~\cite{cady} uses NLP to parse datasheets into a formal rule
database and checks uploaded netlist and component list files against it.
Cadstrom~\cite{cadstrom} combines first-principles physics simulation
with generative AI to validate schematics and PCB layouts. Both are
standalone tools with no version control, collaboration, or CI/CD
integration. DRCY differs in that it operates directly on native
schematic representations within an integrated version control and
collaboration platform, uses agentic LLM reasoning with autonomous
datasheet retrieval, and is embedded in a CI/CD pipeline that triggers
automatically on design review submissions.

Recent work has also investigated applying LLMs to hardware
\emph{generation}. Chip-Chat
\cite{chipchat} and RTLCoder~\cite{rtlcoder} target Verilog synthesis, while
surveys~\cite{llm_eda_acm,llm_eda_survey} map the broader
LLM-for-EDA landscape. DRCY addresses the complementary problem of hardware
\emph{verification}, operating on completed designs rather than generating new
ones.

DRCY's architecture draws on the compound AI systems paradigm~\cite{compound},
where multiple LLM calls are composed with tool use and self-evaluation. The
multi-run consensus mechanism relates to self-consistency
prompting~\cite{selfconsistency}, adapted here for structured engineering
analysis rather than chain-of-thought reasoning. LLM-based code review tools~\cite{li2022codereviewer} for software are
conceptually related, but hardware review requires domain-specific capabilities
such as datasheet retrieval, pin-level analysis, and netlist reasoning that have no
direct software analog.

To our knowledge, DRCY is the first system to combine autonomous datasheet
retrieval with LLM-based connection analysis in a production CI/CD
environment for PCB design review.

%% ============================================================
\section{Conclusion}

In this paper, we introduced the design of DRCY, the first
production-deployed multi-agent LLM system for automated hardware design
reviews. DRCY decomposes a design review task into a comprehensive
multi-agent pipeline in a way that increases accuracy and prevents
context degradation. It takes advantage of AllSpice Hub, which is a
platform already built for collaborative hardware design. By using a
custom agent architecture built specifically for the needs of schematic
designs, DRCY provides hardware design teams with actionable,
evidence-based feedback integrated directly into their design review
workflow. The system is in active production use and available as a CI/CD
action on AllSpice Hub, where it is actively used by hardware teams
across multiple industries.

%% ============================================================
\bibliographystyle{ACM-Reference-Format}
\bibliography{references}

@misc{allspice,
  author       = {{AllSpice}},
  title        = {Allspice Platform},
  year         = {2026},
  url          = {https://www.allspice.io/product},
}

@misc{cadence_erc,
  author       = {{Cadence Design Systems}},
  title        = {OrCAD Electrical Rules Check},
  year         = {2026},
  url          = {https://www.ema-eda.com/products/cadence-orcad/orcad-sigrity-erc},
}

@misc{cadence_drc,
  author = {{Cadence Design Systems}},
  title = {OrCAD Design Rule Check},
  year = {2026},
  url = {https://www.ema-eda.com/courses/orcad-capture-17-4-2021-walk-through/lessons/capture-walk-through-9-design-rule-check-2/}
}

@misc{altium_erc,
  author       = {{Altium}},
  title        = {{Altium Designer} Electrical Rules Check},
  year         = {2026},
  url          = {https://www.altium.com/documentation/altium-circuitmaker/electrical-rules-check},
}

@misc{altium_overview,
  author       = {{Altium}},
  title        = {Altium Designer},
  url = {https://www.altium.com/altium-designer},
  year         = {2026}
}

@misc{cadence_overview,
  author = {{Cadence Design Systems}},
  title = {Cadence OrCAD},
  year = {2026},
  url = {https://www.cadence.com/en_US/home/tools/pcb-design-and-analysis/orcad.html}
}

@misc{kicad_overview,
  author       = {{KiCad}},
  title        = {KiCad},
  year         = {2026},
  url          = {https://www.kicad.org/},
}

@misc{altium_drc,
  author = {{Altium}},
  title = {{Altium Designer} Design Rules Check},
  year = {2026},
  url = {https://www.altium.com/documentation/altium-designer/pcb/drc}
}

@book{art_of_electronics,
  author    = {Horowitz, Paul and Hill, Winfield},
  title     = {The Art of Electronics},
  edition   = {3rd},
  publisher = {Cambridge University Press},
  year      = {2015},
  isbn      = {978-0521809269},
}

@techreport{ipc2221,
  author      = {{IPC}},
  title       = {{IPC-2221B}: Generic Standard on Printed Board Design},
  institution = {IPC --- Association Connecting Electronics Industries},
  year        = {2012},
}

@article{llm_eda_acm,
  author  = {He, Zhuolun and Pu, Yuan and Wu, Haoyuan and Qiu, Tairu and Yu, Bei},
  title   = {Large Language Models for {EDA}: Future or Mirage?},
  journal = {ACM Transactions on Design Automation of Electronic Systems},
  volume  = {30},
  number  = {6},
  year    = {2025},
  doi     = {10.1145/3736167},
}

@misc{cady,
  author = {{CADY}},
  title  = {CADY: Automatic AI Analysis for Eletrical Schematics},
  year   = {2026},
  url    = {https://cadysolutions.com/},
}

@misc{cadstrom,
  author = {{Cadstrom}},
  title  = {Cadstrom: Catch {PCB} Errors Before They Cost You Weeks},
  year   = {2026},
  url    = {https://www.cadstrom.io/},
}

@inproceedings{chipchat,
  author    = {Blocklove, Jason and Garg, Siddharth and Karri, Ramesh and Pearce, Hammond},
  title     = {{Chip-Chat}: Challenges and Opportunities in Conversational Hardware Design},
  booktitle = {Proceedings of the 2023 ACM/IEEE 5th Workshop on Machine Learning for CAD (MLCAD)},
  year      = {2023},
  doi       = {10.1109/MLCAD58807.2023.10299874},
}

@article{rtlcoder,
  author  = {Liu, Shang and Fang, Wenji and Lu, Yao and Zhang, Qijun and Zhang, Hongce and Xie, Zhiyao},
  title   = {{RTLCoder}: Outperforming {GPT-3.5} in Design {RTL} Generation with Our Open-Source Dataset and Lightweight Solution},
  journal = {IEEE Transactions on Computer-Aided Design of Integrated Circuits and Systems},
  year    = {2024},
  doi     = {10.1109/TCAD.2024.3431278},
}

@article{llm_eda_survey,
  title={Llm4eda: Emerging progress in large language models for electronic design automation},
  author={Zhong, Ruizhe and Du, Xingbo and Kai, Shixiong and Tang, Zhentao and Xu, Siyuan and Zhen, Hui-Ling and Hao, Jianye and Xu, Qiang and Yuan, Mingxuan and Yan, Junchi},
  journal={arXiv preprint arXiv:2401.12224},
  year={2023}
}

@article{compound,
  author  = {Zaharia, Matei and Khattab, Omar and Chen, Lingjiao and Goldstein, Jared and Liang, Percy and Beutel, Alex and Stoica, Ion and Re, Christopher and others},
  title   = {The Shift from Models to Compound {AI} Systems},
  journal = {Berkeley AI Research Blog},
  year    = {2024},
  url     = {https://bair.berkeley.edu/blog/2024/02/18/compound-ai-systems/},
}

@inproceedings{selfconsistency,
  author    = {Wang, Xuezhi and Wei, Jason and Schuurmans, Dale and Le, Quoc and Chi, Ed and Narang, Sharan and Chowdhery, Aakanksha and Zhou, Denny},
  title     = {Self-Consistency Improves Chain of Thought Reasoning in Language Models},
  booktitle = {Proceedings of the 11th International Conference on Learning Representations (ICLR)},
  year      = {2023},
}

@inproceedings{li2022codereviewer,
  author = {Li, Zhiyu and Lu, Shuai and Guo, Daya and Duan, Nan and Jannu, Shailesh and Jenks, Grant and Majumder, Deep and Green, Jared and Svyatkovskiy, Alexey and Fu, Shengyu and Sundaresan, Neel},
  title     = {Automating code review activities by large-scale pre-training},
  booktitle = {Proceedings of the 30th ACM Joint European Software Engineering Conference and Symposium on the Foundations of Software Engineering},
  year      = {2022},
  doi = {10.1145/3540250.3549081},
}

\end{document}